\documentclass[aps,prl,superscriptaddress,twocolumn]{revtex4-2}
\usepackage{amsfonts,amssymb,amsmath,amsthm,mathrsfs,graphicx,bm,color,multirow}
\usepackage{subfigure}
\usepackage{dcolumn}
\usepackage[usenames,dvipsnames]{xcolor}
\usepackage[all]{xy}
\usepackage{comment}
\usepackage[colorlinks=true,allcolors=blue]{hyperref}

\renewcommand{\bf}[1]{\textnormal{\textbf{#1}}}
\newcommand{\BZ}{\mathrm{BZ}}
\newcommand{\tr}{\mathrm{Tr}}

\newcommand{\ket}[1]{| #1 \rangle}
\newcommand{\bra}[1]{\langle #1|}

\begin{document}

\author{Carolina Paiva}

\affiliation{School of Physics and Astronomy, Tel Aviv University, Tel Aviv 6997801, Israel}

\author{Jie Wang}

\affiliation{Department of Physics, Temple University, Philadelphia, Pennsylvania, 19122, USA}

\author{Tomoki Ozawa}

\affiliation{Advanced Institute for Materials Research (WPI-AIMR), Tohoku University, Sendai 980-8577, Japan}
\author{Bruno Mera}

\affiliation{Instituto de Telecomunica\c{c}\~oes and Departmento de Matem\'{a}tica, Instituto Superior T\'ecnico, Universidade de Lisboa, Avenida Rovisco Pais 1, 1049-001 Lisboa, Portugal}

\affiliation{Advanced Institute for Materials Research (WPI-AIMR), Tohoku University, Sendai 980-8577, Japan}
\title{Geometrical Responses of Generalized Landau Levels: Structure Factor and the Quantized Hall Viscosity}
%
%
\begin{abstract}
We present a new geometric characterization of generalized Landau levels (GLLs). The GLLs are a generalization of Landau levels to non-uniform Berry curvature, and are mathematically defined in terms of a holomorphic curve---an ideal K\"ahler band---and its associated unitary Frenet-Serret moving frame.  Here, we find that GLLs are harmonic maps from the Brillouin zone to the complex projective space and they are critical points of the Dirichlet energy functional, as well as the static structure factor up to fourth order. We also find that filled GLLs exhibit quantized Hall viscosity, similar to the ordinary Landau levels. These results establish GLLs as a versatile generalization of Landau levels. 
\end{abstract}
\maketitle 
\emph{Introduction}---
Quantum Hall states are distinguished by their precise quantization of transport coefficients, such as the Hall conductivity~\cite{TKNN:1995}, which reflects the system’s topological invariants. Beyond conductivity, additional transport coefficients have been identified that provide deeper insights into the interplay between topology and geometry. Among these, the Hall viscosity has emerged as a key geometric transport coefficient, capturing the response of quantum Hall states under adiabatic changes to the system’s metric~\cite{avron:seiler:zograf:95,Read:2011,Haldane:2011}. 
In two-dimensional systems, such metric deformations, provided the area is kept constant, are equivalent to changes in the complex structure, which, for the torus, is parametrized by the modular parameter $\tau=\tau_1+i\tau_2$ with $\tau\in\mathbb{H}$ and $\mathbb{H}$ the upper half plane. The Hall viscosity can thus be understood as a Berry curvature on the moduli space of complex structures, which governs the response of the quantum Hall state to adiabatic deformations of $\tau$. This connection was first established in the seminal work of Avron, Seiler, and Zograf~\cite{avron:seiler:zograf:95}, tying it to the intrinsic geometry of the quantum Hall state. Importantly, the corresponding dissipationless transport coefficient, $\eta_{H}$, is quantized and determined by the 1st Chern number associated with this curvature~\cite{klevtsov:wiegmann:2015}. This insight not only highlighted Hall viscosity as an important feature of two-dimensional gapped systems that break time-reversal symmetry but also positioned it as a fundamental topological invariant, complementing the Hall conductivity. In~\cite{klevtsov:wiegmann:2015} an extension of the concept of geometric adiabatic transport is done to surfaces of higher genus ($g>1$) and introduces a novel transport coefficient, the central charge~\cite{Bradlyn:2015,Klevtsov:2016}, which arises from the gravitational anomaly. This central charge quantifies the universal response of quantum Hall states to geometric deformations, linking it to topological and conformal field theory invariants.

Recently the concept of generalized Landau levels (GLLs) has emerged as a systematic extension of Landau levels~\cite{liu:mera:fujimoto:ozawa:wang:24,fujimoto2024higher}. Unlike standard Landau levels, GLLs incorporate non-uniform quantum geometric while preserving the quantized value of the trace of the quantum metric~\cite{liu:mera:fujimoto:ozawa:wang:24}.
The lowest GLLs are defined as bands that saturate the trace geometric bound~\cite{rahul:14,claassen:15} and formally corresponds to holomorphic curves~\cite{ozawa:mera:2021,mera:ozawa:21,liu:mera:fujimoto:ozawa:wang:24}. These bands have implications for abelian fractional Chern insulator in moiré flatbands~\cite{jie:cano:millis:liu:yang:21,ledwith:tarnopolsky:khalaf:vishwanath:20,ledwith:vishwanath:parker:22,jie:liu:22,ledwith:vishwanath:khalaf:22}. The higher Landau level generalizations are formulated in terms of the Frenet-Serret moving frames that one can attach to the holomorphic curves. GLLs exhibit a quantized integrated trace of the quantum metric, $\frac{1}{2\pi}\int_{\BZ^2}\!\mathrm{tr}(g)d^2\bf{k}= (2n+1)|\mathcal{C}|$, and have implications for non-Abelian topological orders~\cite{liu:mera:fujimoto:ozawa:wang:24,fujimoto2024higher,nonabelian:di:24,nonabelian:cho:24,nonabelian:zhang:24,nonabelian:fu:24}.
However, despite the recent importance of these bands, a comprehensive understanding of their transport coefficients remains lacking. 
In particular, it is crucial to investigate how inhomogeneity in the Berry curvature influences these coefficients and their quantization, shedding light on the geometric and topological responses of GLLs. In this paper, we establish a comprehensive understanding of GLLs by identifying them as harmonic maps from the Brillouin zone to the complex projective space. Because these harmonic maps depend explicitly on a choice of a conformal class of a flat metric on the Brillouin zone or, equivalently, on a modular parameter $\tau$, it is possible to define the viscosity in the same way as in the work of Avron \emph{et al}~\cite{avron:seiler:zograf:95}. We compute the Hall viscosity for fully filled harmonic bands and demonstrate that the quantization of the transport coefficient as $2n+1$ for the $n$th GLL—previously established for filled Landau levels—remains valid. This result not only gives a physical meaning to the $2n+1$ result of the integrated trace of the quantum metric for GLLs~\cite{liu:mera:fujimoto:ozawa:wang:24}, but also unifies the geometric and topological response properties of GLLs with those of conventional Landau levels.

\emph{Harmonic maps and generalized Landau levels}---
Suppose we have an isolated, nondegenerate band described by an orthogonal rank $1$ projector $P(\bf{k})$ in $\mathbb{C}^N$, $N$ being the total number of bands. Assume this band is completely filled. The static structure factor is defined as the equal time density-density correlation function
\begin{align}
S(\bf{q}) &:=\frac{\langle \rho(-\bf{q})\rho(\bf{q})\rangle_c}{V} =\frac{1}{2}\int_{\BZ^2}\!\! \tr\left[P(\bf{k})\!-\! P(\bf{k}+\bf{q})\right]^2 d^2\bf{k}.
\label{eq:structure_factor_projectors}
\end{align}
Here, $\rho(\bf{q})$ is the Fourier transform of the charge density $\rho(\bf{r})$ at transferred momentum $\bf{q}$, $V$ the volume, and $\langle\cdot\rangle$ denotes the average on the quantum state obtained by filling the band. Additionally, $\tr$ denotes the trace over the internal degrees of freedom, not to be confused with $\mathrm{tr}$ which we introduce below.
The second equality follows from Wick's theorem and the fact that $P(\bf{k})$ is an orthogonal projector; see Ref.~\cite{onishi:fu:24}. In the limit $\bf{q}\rightarrow 0$ the first nontrivial contribution to the expansion of Eq.~\eqref{eq:structure_factor_projectors} arises from the second-order term. This term is described by the quantum metric $g=(1/2)\tr \left(dPdP\right)=g_{ij}dk^idk^j$. The corresponding contribution to the structure factor, as first shown in~\cite{onishi:fu:24}, is:
\begin{align}
S_2(\bf{q}):=\int_{\BZ^2}d^2\bf{k}\; g_{ij} q^i q^j=\widetilde{g}_{ij}q^iq^j.  \end{align}
Where $\bf{q}$ is identified as a vector field in the Brillouin zone $BZ^2$ that generates translations, and $\widetilde{g}$ is the twist-angle space quantum metric. This flat metric, in the thermodynamic limit, is obtained by averaging the quantum metric with respect to translations in $\BZ^2$, i.e. $\widetilde{g}_{ij}=\int_{\BZ^2}g_{ij}\; d^2\bf{k}$. Throughout this discussion, we adopt the Einstein summation convention. Translation-invariant, uniform complex structures on the $\BZ^2$ are described by a complex coordinate $z=-k^y+\tau k^x$, with modular parameter $\tau\in \mathbb{H}$. By varying $\tau$, we can describe any translation-invariant complex structure. The vector field $\bf{q}$ can equivalently be expressed using the complex coordinate $q=-q^y+\tau q^x$. The structure factor $S_2(\bf{q})$ can then be written as
\begin{align}
S_2(\bf{q})=\widetilde{g}_{zz}q^2 +2\widetilde{g}_{\bar{z} z}\bar{q}q +\widetilde{g}_{\bar{z}\bar{z}}\bar{q}^2,
\end{align}
Since $\widetilde{g}$ is a flat metric, there exists a translation-invariant complex structure, described by some modular parameter $\tau$, for which the anomalous terms ($q^2$ and $\bar{q}^2$) vanish. We now restrict the bands of interest to those with this fixed valued of $\tau$. Physically, $\tau$ describes some degree of anisotropy in the localization properties of the band insulator~\cite{mera:2020}. The quantity determining $S_2(\bf{q})$ is then $\widetilde{g}_{\bar{z}z}$ for this specific complex structure. It can be computed as:
\begin{align}
\widetilde{g}_{\bar{z}z} &=\int_{\BZ^2} g_{\bar{z}z} d^2\bf{k}=\frac{1}{2}\int_{\BZ^2}  h^{ij}g_{ij}\; \sqrt{\det(h)}d^2\bf{k}
\nonumber \\
&=\frac{1}{4}\int_{\BZ^2} h^{ij}\tr\left(\frac{\partial P}{\partial k^i}\frac{\partial P}{\partial k^j}\right)\sqrt{\det(h)} d^2\bf{k}=:E(P),
\end{align}
where $h^{ij}$ are the components of the inverse of the metric $h=h_{z\bar{z}}|dz|^2$, where $h_{z\bar{z}}$ is some smooth periodic function on $\BZ^2$. The quantity $E(P)$ is the Dirichlet energy functional for the map $P:\BZ^2\to \mathbb{C}P^{N-1}$ determined by the rank $1$ projector $P(\bf{k})$. We note that, since the Brillouin zone is two-dimensional, $E(P)$ does not depend on how we choose the smooth periodic function $h_{z\bar{z}}$. 
Therefore we can assume, without loss of generality, that $h$ is a unimodular flat metric, i.e., $\sqrt{\det(h)}=h_{z\bar{z}}=1$, reducing $E(P)$ to the so-called \emph{integrated trace} $\frac{1}{2}\int_{\BZ^2}\mathrm{tr}(g)d^2\bf{k}$, with $\mathrm{tr}(g)=h^{ij}g_{ij}$. This quantity has been understood as a signature of Landau-level mimicry, which is no coincidence. As we will argue below, this connection arises by relating the critical points of $E(P)$ to GLLs, as introduced in Ref.~\cite{liu:mera:fujimoto:ozawa:wang:24}. We remark that $E(P)$ also corresponds to the Polyakov action functional (in Euclidean signature) for a string propagating in the background ``spacetime'' $\mathbb{C}P^{N-1}$, with the worldsheet given by $\BZ^2$. 

The critical points of $E(P)$---which, for small $\bf{q}$, also correspond to critical points of the structure factor $S(\bf{q})$ as a functional of $P$---are known as \emph{harmonic maps}. Under the assumption that the Chern number $\mathcal{C}\neq 0$, these maps have been classified in the works of Eels and Wood in~\cite{eells:wood:83}. Remarkably, this classification is entirely captured by the GLL construction in \cite{liu:mera:fujimoto:ozawa:wang:24}. Specifically, each GLL defines a (full \footnote{The band is such that the image of the associated map to projective space is not contained in any proper linear subspace. This condition can always be met by restricting the total number of bands in the system.}) harmonic map to projective space, and all (full) harmonic maps can be realized in this manner. 
Recently, in Ref.~\cite{onishi:avdoshkin:fu:2024}, the term \emph{harmonic bands} was used to define bands that saturate a bound on the fourth order term of the small $\bf{q}$ expansion of the structure factor $S(\bf{q})$, as these bands satisfy an analog of Laplace's equation. We show that the harmonic bands defined in Ref.~\cite{onishi:avdoshkin:fu:2024}, restricted to two dimensional setups and nontrivial Chern number $\mathcal{C}\neq 0$, are completely equivalent to the GLLs.

In the construction of GLLs, as detailed in Ref.~\cite{liu:mera:fujimoto:ozawa:wang:24}, one begins by fixing an ideal K\"ahler band, which serves as the zeroth GLL. Mathematically, this band is described by a holomorphic curve $P:=P_0:\BZ^2\to\mathbb{C}P^{N-1}$, represented by a holomorphic vector $u=u(z)\in\mathbb{C}^N$, with $z=-k^y+\tau k^x$. To construct the higher GLLs, one collects the holomorphic derivatives $u^{(k)}(z)$ for $k=0,\dots, N-1$ and performs the Gram-Schmidt orthogonalization procedure. This process generates a unitary \emph{Frenet-frame} of bands, $U=[u_0,\dots, u_{N-1}]$. As shown in Ref.~\cite{liu:mera:fujimoto:ozawa:wang:24}, the Frenet frame is unique up to a phase, and its Maurer-Cartan one-form $\theta=U^{-1}dU$ is tridiagonal:
\begin{align}
du_i = u_{i-1}\theta^{i-1}_{\ i}+u_i \theta^{i}_{\; i} + u_{i+1} \theta^{i+1}_{\; i},
\label{eq: Frenet-Serret equations}
\end{align}
with $\theta^{i+1}_{\; i}=\langle u_{i+1}|\frac{\partial}{\partial z}|u_i\rangle dz$, $\theta^{i-1}_{\; i}=\langle u_{i-1}|\frac{\partial}{\partial \bar{z}}|u_i\rangle d\bar{z}$, and $\theta^i_{\; i}=\langle u_i|d|u_i\rangle$ corresponds to the Berry connection of the $i$-th GLL.

Remarkably, as we show in the Supplemental Material (SM), each $P_i:=\ket{u_i}\bra{u_i}$, with $i=0,...,N-1$, defines a harmonic map, and hence it is a critical point of $E(P)$. In particular $P=P_0$, the holomorphic map representing the ideal K\"ahler band, is also a harmonic map. Varying $E(P)$ yields, after integration by parts:
\begin{align}
\delta E= -\frac{1}{2}\int_{\BZ^2} \tr\left[\delta P\frac{\partial^2 P}{\partial z \partial \bar{z}}\right]d^2\bf{k}.
\end{align}
Given that $P^2=P$, variations of $P$ satisfy $
\delta P = Q\delta P P + P\delta P Q$,
where $Q=I-P$ is the orthogonal complement projector. This allows us to write:
\begin{align}
\delta E= -\frac{1}{2}\int_{\BZ^2}\tr\left[\delta P\left(Q\frac{\partial^2 P}{\partial z \partial \bar{z}}P +P\frac{\partial^2 P}{\partial z \partial \bar{z}}Q\right)\right].
\end{align}
The critical points are determined by setting the integrand to zero, yielding the only independent equation \footnote{the vanishing of the sum in the integrand, due to orthogonality of $P$ and $Q$, implies the vanishing of each of them and, also, they are the Hermitian conjugate of each other.}
\begin{align}
Q\frac{\partial^2 P}{\partial z\partial \bar{z}} P=0,
\label{eq: harmonic band projector equation}
\end{align}
which is precisely the result obtained in Ref.~\cite{onishi:avdoshkin:fu:2024}, for the case of the isotropic complex structure $\tau=i$. This confirms that the GLLs studied here align with those defined in Ref.~\cite{onishi:avdoshkin:fu:2024}, while also accommodating for more general complex structures. 

For completeness, we demonstrate in the SM that all $P_i$'s are critical points of $E$, and thus correspond to harmonic maps. The key elements of the proof are the Frenet-Serret equations Eq.~\eqref{eq: Frenet-Serret equations} and the structure equation satisfied by the Maurer-Cartan one-form $d\theta+\theta\wedge \theta=0$.

\emph{Geometric response GLL}---
We now turn to the response of GLLs, to shear strain, specifically computing the Hall viscosity associated with a completely filled GLL. We show that the Hall viscosity is quantized to $2n+1$, where $n$ is the GLL index, just like what happens in the case of ordinary Landau levels, further stressing how GLLs mimic the ordinary Landau levels.

The harmonic maps considered here depend, as previously explained, on the choice of a conformal class of a flat metric on the Brillouin zone, which can equivalently be expressed in terms of the modular parameter $\tau$. This framework enables the definition of the Hall viscosity of GLLs as the Berry curvature in the moduli space of complex structures. Crucially, this approach allows for a consistent definition of Hall viscosity even in the absence of a Hamiltonian explicitly dependent on the metric. Moreover, this formulation faithfully reproduces the known results for ordinary Landau levels.

The proof strategy is as follows. We first show that the fully filled the $N$ first GLL is a many-body state that behaves like an ideal K\"ahler band in twist-angle space, with uniform quantum geometry. Calabi's rigidity theorem then implies that these states must differ at most by a constant unitary. The Hall viscosity of all GLLs can then be derived from additivity of Berry curvature with respect to filling additional bands.

For simplicity, we focus on ideal K\"ahler bands with Chern number $\mathcal{C}=1$~\cite{liu:mera:fujimoto:ozawa:wang:24}. The arguments presented generalize to $\mathcal{C}> 0$.

Let $\ket{u_{\bf{k}}}$ denote an ideal K\"ahler band Bloch wavefunction in a holomorphic gauge, where $\frac{\partial}{\partial \bar z}\ket{u_{\bf{k}}}=0$. The $N$th Slater determinant state, $\ket{\Psi_{N,\bf{k}}}$, corresponding to filling the first associated $N$ GLLs, forms a K\"ahler band and is represented by a holomorphic state vector
\begin{align}
\ket{\Psi_{N,\bf{k}}}= \ket{u_{\bf{k}}}\wedge \frac{\partial}{\partial z}\ket{u_{\bf{k}}}\wedge \dots \wedge   \frac{\partial^{N-1}}{\partial z^{N-1}}\ket{u_{\bf{k}}},\ N=1,2,\dots  
\end{align}
Since $\ket{u_{\bf{k}}}$ can be expressed as a holomorphic function of $z=-k^y+\tau k^x$ and $\tau$ via theta functions, $\ket{\Psi_{N\bf{k}}}$ is also holomorphic in both variables. To analyze the many-body state of the system, we place it on a discrete torus with $L$ sites in both real space directions and impose twisted boundary conditions parameterized by twist angles $\bm{\theta}=(\theta^1,\theta^2)$. These angles specify the allowed momenta in each direction, taking the form $\bf{k}=\bf{m}/L +\bm{\theta}/L$, where $\bf{m}=(m^x,n^y)$ with $m^i\in\{0,\dots, L-1\}$ and $i=x,y$. Note that we assume $\BZ^2=\mathbb{R}^2/\mathbb{Z}^2$, implying that $k^x$ and $k^y$ in the standard fundamental domain take values in $[0,1)$. The many-body state corresponding to filling the first $N$ GLLs on this discrete torus is given by the Slater determinant:
\begin{align*}
\ket{\widetilde{\Psi}_{N,\bm{\theta}}}=\bigwedge_{\bf{m}}\ket{\Psi_{N, \frac{\bf{m}}{L}+\frac{\bm{\theta}}{L}}}    
\end{align*}
By construction, since $\ket{\Psi_{N,\bf{k}}}$ is holomorphic in $z=-k^y+\tau k^x$, it follows that $\ket{\widetilde{\Psi}_{N,\bm{\theta}}}$ is holomorphic in $\theta= -\theta^y+ \tau \theta^x$. In the thermodynamic limit $L\to\infty$, the twist-angle space quantum geometry becomes uniform, exhibiting translation invariance in the $\theta$ variable. We now make use of Calabi's rigidity theorem applied to the case of holomorphic curves on projective space, a proof of which is presented in the SM. Here, $\ket{\widetilde{\Psi}_{N,\bm{\theta}}}$ defines a holomorphic map to projective space with Kähler form
\begin{align}
\widetilde{\omega}_N= \frac{i}{2}\frac{\pi N}{\tau_2} d\theta\wedge d\bar\theta.
\label{eq: Kaehler form for widetilde Psi N}
\end{align}
The above follows from translation-invariance in $\theta$, together with the fact that filling the first $N$ GLLs gives a band with Chern number $N$, which then fixes the integral of the K\"ahler form, see~\cite{liu:mera:fujimoto:ozawa:wang:24}.
We now observe that the Kähler form derived above is associated to a holomorphic map to a projective space, known as the color-entangled wavefunction, which describes a uniform quantum geometry ideal Kähler band, with the momentum variable replaced by the twist-angle. 
According to Calabi's rigidity theorem~\cite{calabi:53}, $\ket{\widetilde{\Psi}_{N,\bm{\theta}}}$ differs from the color-entangled wavefunction, up to a multiplicative nonvanishing holomorphic function, only by a constant unitary operator, which does not alter the quantum geometry. Remarkably, for $N=1$, the color-entangled wavefunction is just the lowest Landau level Bloch wavefunction and, hence, up to a constant unitary and with the replacement $\bf{k}\to\bm{\theta}$, so is $\ket{\widetilde{\Psi}_{1,\bm{\theta}}}=:\ket{\widetilde{u}_{\bm{\theta}}}$. 

The quantum metric associated with $\ket{\widetilde{\Psi}_{N,\bm{\theta}}}$ is denoted by $\widetilde{\gamma}_N = \widetilde{h}_N |dz|^2$, where $\widetilde{h}_N=\frac{\partial^2 \log\bra{\widetilde{\Psi}_{N,\bm{\theta}}}\widetilde{\Psi}_{N,\bm{\theta}}\rangle}{\partial \theta \partial \bar \theta}$ and in the thermodynamic limit $\widetilde{h}_N=\frac{\pi N}{\tau_2}$, \emph{cf.} Eq.\eqref{eq: Kaehler form for widetilde Psi N} for the associated compatible K\"ahler form. Consider the unitary frame of GLLs, $\ket{u_{0,\bf{k}}}, \ket{u_{1,\bf{k}}}, \dots$, obtained via Gram-Schmidt orthogonalization of $\frac{\partial^n}{\partial z^n}\ket{u_{\bf{k}}},\ n=0,1,\dots$, as described in Ref.~\cite{liu:mera:fujimoto:ozawa:wang:24}. By construction, the $\ket{u_{n,\bf{k}}}$ states inherit $\tau$ dependence. Consequently, the adiabatic curvature determining the Hall viscosity of the state:
\begin{align*}
\ket{\widetilde{u}_{n,\bm{\theta}}}=\bigwedge_{\bf{m}}\ket{ u_{n, \frac{\bf{m}}{N}+\frac{\bm{\theta}}{N}}}
\end{align*}
is well-defined and given by:
\begin{align}
\eta_{n}(\bm{\theta})=-i\bra{d \widetilde{u}_{n,\bm{\theta}}} d\widetilde{u}_{n,\bm{\theta}}\rangle,
\end{align}
where $d=\partial+\overline{\partial}$, and $\partial=d\tau\frac{\partial}{\partial\tau}$.

In the thermodynamic limit, we claim that $\ket{\widetilde{u}_{n,\bm{\theta}}}$ is, from the point of view of quantum geometry, indistinguishable from the many-body wavefunction of the filled $n$th Landau level. Moreover, due to the quantum geometry being uniform, it is also indistinguishable from the Bloch wavefunction of the filled $n$th Landau level with the replacement $\bf{k}\to\bm{\theta}$. First observe that the Kähler form associated with filling the first $N$ Landau levels, $\widetilde{\omega}_N$, uniquely determines the state $\ket{\widetilde{\Psi}_{N,\bm{\theta}}}$ (up to a constant unitary) via Calabi's rigidity theorem. Transitioning from filling $N$ to $N+1$ Landau levels corresponds precisely to wedging with the $n$th Landau level. Thus, the $\ket{\widetilde{\Psi}_{N,\bm{\theta}}}$'s, up to a constant unitary and multiplication by a nonvanishing holomorphic function of $\theta$, reproduce, in the variable $\bm{\theta}$, the Slater determinant states as in the GLL construction, with the starting K\"ahler band the lowest Landau level itself. In the thermodynamic limit and up to constant unitary and multiplication my holomorphic nonvanishing function of $\theta$, the many-body state $\ket{\widetilde{\Psi}_{N,\bm{\theta}}}$ can then be expressed as:
\begin{align}
\ket{\widetilde{\Psi}_{N,\bm{\theta}}}=\ket{\widetilde{u}_{\bm{\theta}}}\wedge \frac{\partial}{\partial \theta}\ket{\widetilde{u}_{\bm{\theta}}}\wedge \dots \wedge \frac{\partial^{N-1}}{\partial \theta^{N-1}}\ket{\widetilde{u}_{\bm{\theta}}}, \ N=1,2,\dots,
\end{align}
where we can intepret $\ket{\widetilde{u}_{\bm{\theta}}}$ as the lowest Landau level Bloch wavefunction, with the replacement $\bf{k}\to \bm{\theta}$. Note that the above formula is in perfect analogy with the local momentum-space one, but this one describes the macroscopic many-body states obtained by filling GLLs up to level $N$. A key observation is that the Berry curvature of a Slater determinant state of orthogonal states is the sum of contributions from each element in the determinant. Since $\ket{\widetilde{\Psi}_{N,\bm{\theta}}}=\pm||\widetilde{\Psi}_{N,\bm{\theta}}||\left( \ket{\widetilde{u}_{0,\bm{\theta}}}\wedge \dots \wedge \ket{\widetilde{u}_{N-1,\bm{\theta}}}\right)$ and $\ket{\widetilde{\Psi}_{N,\bm{\theta}}}$ is holomorphic in $\tau$, the Berry curvature in a holomorphic gauge is completely determined by the normalization of $\ket{\widetilde{\Psi}_{N,\bm{\theta}}}$. It is given by $i\partial\overline{\partial}\log \bra{\widetilde{\Psi}_{N,\bm{\theta}}}\widetilde{\Psi}_{N,\bm{\theta}}\rangle$. Furthermore, because $\ket{\widetilde{\Psi}_{N+1,\bm{\theta}}}\propto \ket{\widetilde{u}_{n,\bm{\theta}}}\wedge \ket{\widetilde{\Psi}_{N,\bm{\theta}}}$, it follows that the viscosity of the filled $n$th GLL is:
\begin{align}
\eta_{n}(\bm{\theta}) =i\partial \overline{\partial} \log \bra{\widetilde{\Psi}_{n+1,\bm{\theta}}}\widetilde{\Psi}_{n+1,\bm{\theta}}\rangle-i\partial \overline{\partial} \log \bra{\widetilde{\Psi}_{n,\bm{\theta}}}\widetilde{\Psi}_{n,\bm{\theta}}\rangle 
\label{eq:etan}
\end{align}
Here, we have set $N=n$. Even though we initially distinguished between $N$ and $n$, it is valid to equate them since $\ket{\widetilde{\Psi}_{0\bm{\theta}}}$ is a non-existing state with normalization 1, implying zero curvature. From the above formula, it becomes evident that $\ket{\widetilde{u}_{n,\bm{\theta}}}$ is indistinguishable, from the quantum geometry point of view from filling the $n$th LL. Due to Calabi's rigidity theorem the right-hand side is indistinguishable: $\ket{\widetilde{\Psi}_{n+1,\bm{\theta}}}$ and $\ket{\widetilde{\Psi}_{n,\bm{\theta}}}$ differ from the many-body states corresponding to filling the first $n+1$ and $n$ Landau levels, respectively, only by a $\bm{\theta}$-independent unitary transformation. This transformation does not affect the normalizations and thus leaves the quantum geometric properties unchanged. Note that a similiar formula to Eq.~\eqref{eq:etan} holds for the Berry curvature in twist-angle space, i.e. taking derivatives with respect to the variable $\theta$ instead of $\tau$.
Using the result:
\begin{align}
\widetilde{h}_n= \frac{\bra{\widetilde{\Psi}_{n+1,\bm{\theta}}}\widetilde{\Psi}_{n+1,\bm{\theta}}\rangle\bra{\widetilde{\Psi}_{n-1,\bm{\theta}}}\widetilde{\Psi}_{n-1,\bm{\theta}}\rangle}{\bra{\widetilde{\Psi}_{n,\bm{\theta}}}\widetilde{\Psi}_{n,\bm{\theta}}\rangle^2},
\label{eq:metricandnormalizations}
\end{align}
proved in the Supplemental Material (SM), we derive the recursion relation for the viscosities of filled GLLs $\eta_n$'s,
\begin{align}
\eta_{n}(\bm{\theta})-\eta_{n-1}(\bm{\theta})=i\partial\overline{\partial}\log \widetilde{h}_n.   
\end{align}
In the thermodynamic limit $L\to\infty$, we know that $\widetilde{h}_n= \frac{\pi n}{\tau_2}$, so:
\begin{align}
\eta_{n}(\bm{\theta})-\eta_{n-1}(\bm{\theta})=i\partial\bar{\partial}\log\left( \frac{\pi n}{\tau_2}\right)=\frac{i}{4}\frac{1}{\tau_2^2}d\tau\wedge d\bar{\tau}.
\end{align}
For $\eta_0$, the viscosity of the ideal Kähler band with $\bra{\widetilde{\Psi}_{1,\bm{\theta}}}\widetilde{\Psi}_{1,\bm{\theta}}\rangle=(1/\sqrt{2\tau_2})e^{-2\pi\tau_2(\theta^2)^2}$, we find: 
\begin{align}
\eta_0(\bm{\theta})= i\partial \overline{\partial}\log \left(\frac{1}{\sqrt{2\tau_2}}\right)= \frac{i}{8}\frac{1}{\tau_2^2}d\tau\wedge d\bar{\tau}.
\end{align}
Solving the recursion relation yields:
\begin{align}
\eta_{n}(\bm{\theta}) = \frac{i}{8}\left(2n+1\right)\frac{1}{\tau_2^2}d\tau\wedge d\bar{\tau} = 2\pi (2n+1)\mu,
\end{align}
where $\mu =\frac{i}{16\pi }\frac{d\tau \wedge d\bar{\tau}}{\tau_2^2}$ 
is a $\mathrm{GL}(2;\mathbb{R})$-invariant area form in the moduli space $\mathcal{M}$ of complex structures on the torus (twist-angle torus and Brillouin zone, in particular), with area $1/24$ . The topological invariant associated with the Hall viscosity of the $n$th GLL, $\eta_{H,n}$, is the 1st Chern number over the moduli space $\mathcal{M}$~\cite{avron:seiler:zograf:95, klevtsov:lecturenotes:2016, klevtsov:wiegmann:2015}
\begin{align}
\frac{\eta_{H,n}}{6}=\int_{\mathcal{M}}\frac{\eta_n(\bm{\theta})}{2\pi}= \frac{1}{24}\left(2n+1\right).
\end{align}
In the thermodynamic limit, it does not depend on the boundary condition as specified by the twist-angles. This topological invariant was previously computed for the lowest Landau level (LLL)~\cite{avron:seiler:zograf:95,klevtsov:wiegmann:2015} and is recovered here as a special case by setting $n=0$.
Notably, $2n+1$ is the projective invariant determined by the $n$th GLL via its quantum metric $g_n$, and equals, up to a numerical constant, the Dirichlet energy
\begin{align}
\frac{E(P_n)}{2\pi} &\!=\!\frac{1}{2\pi}\!\!\int_{\BZ^2}\!\!\mathrm{tr}(g_n)d^2\bf{k} \!=\!\frac{1}{\pi}\!\!\int_{\BZ^2}\!\! \sqrt{\det(g_n)}d^2\bf{k}\!=\! 2n+1.
\end{align}
The relationship between these two quantities for Landau levels follows directly from the computation of their translation-invariant quantum geometries~\cite{ozawa:mera:2021}. Remarkably, this result holds generally because the many-body quantum state of filling the $n$th GLL is equivalent to filling the standard $n$th Landau level. Consequently, the Hall viscosity is the same and quantized.
Equivalence here means that filling the GLL bands from $0$ to $n-1$, i.e. the many-body state $\ket{\widetilde{\Psi}_n}$, and from $0$ to $n$, i.e. $\ket{\widetilde{\Psi}_{n+1}}$, correspond to states that are related to filling the ordinary Landau level bands from $0$ to $n-1$ and from $0$ to $n$, respectively, via some constant unitaries $U_n$ and $U_{n+1}$ in the many-body Hilbert space. The difference between the first two states corresponds to filling the $n$th GLL, while the difference between the latter two represents filling the $n$th LL. Since quantum geometry is insensitive to these unitaries, the geometric responses of these states are macroscopically equivalent. This is a manifestation of how GLLs generalize Landau levels---they differ at the microscopic level but yield the same topological macroscopic response properties, namely the Hall conductivity and the Hall viscosity.

\emph{Conclusions}--- 
In this work, we build upon 
the concept of GLLs, which are here understood in terms of theory of harmonic maps. The latter are naturally defined as critical points of the Dirichlet energy functional $E(P)$, which, in the Bloch band theory, coincides with the integrated trace of the quantum metric. Remarkably, the critical points of $E(P)$ correspond, at small values of $\bf{q}$, to critical points of the structure factor, $S(\bf{q})$. We have shown that GLLs are harmonic maps and noted that in two dimensions with $\mathcal{C}\neq 0$, the converse is also true---all harmonic maps are GLLs. Furthermore, we have shown that the harmonic bands introduced in Ref.\cite{onishi:avdoshkin:fu:2024} are indeed harmonic maps. Through the Frenet-Serret moving frame and associated geometric equations, GLLs offer a unified framework for understanding harmonic bands and explicitly compute their geometric and topological properties.

Furthermore, we connected the geometric response of harmonic bands to shear strain by explicitly computing the Hall viscosity for filled GLLs. Remarkably, we found that the Hall viscosity for the $n$th GLL is quantized, up to numerical factors, to $2n+1$, echoing the behavior of conventional Landau levels. Our findings reinforce the view that GLLs truly generalize Landau levels meaningfully, preserving key macroscopic response properties such as quantized Hall conductivity and viscosity while offering flexibility in their microscopic realizations.

While the current work relies on Calabi's rigidity theorem to establish the connection between filled GLLs and completely filled Landau levels, and thus an equivalence of their macroscopic response properties, an alternative proof based purely on momentum-space formulas would provide a more direct and potentially insightful perspective. This remains an avenue for future exploration.

\emph{Acknowledgements}--- 
We gratefully acknowledge valuable discussions with L.~Fu, Y.~Onishi, A.~Avdoshkin, B.~Bradlyn and Y.~Hashimoto.

B.~M. acknowledges support from the Security and Quantum Information Group (SQIG) in Instituto de Telecomunica\c{c}\~{o}es, Lisbon. This work is funded by FCT (Funda\c{c}\~{a}o para a Ci\^{e}ncia e a Tecnologia) through national funds FCT I.P. and, when eligible, by COMPETE 2020 FEDER funds, under Award UIDB/50008/2020 and the Scientific Employment Stimulus---Individual Call (CEEC Individual)---2022.05522.CEECIND/CP1716/CT0001, with DOI 10.54499/2022.05522.CEECIND/CP1716/CT0001.
T.~O. acknowledges suport from JSPS KAKENHI Grant Number JP24K00548, JST PRESTO Grant No. JPMJPR2353, JST CREST Grant Number JPMJCR19T1.
J.~W. is supported by Temple university start up funding.
C.~P. is acknowledges suport from the US-Israel Binational Science Foundation (BSF, No.2018226), Jerusalem, Israel, and the ISRAEL SCIENCE FOUNDATION (ISF, grant no. 2307/24, 1077/23 and 1916/23) .
\bibliography{bib.bib}

\begin{thebibliography}{32}%
\makeatletter
\providecommand \@ifxundefined [1]{%
 \@ifx{#1\undefined}
}%
\providecommand \@ifnum [1]{%
 \ifnum #1\expandafter \@firstoftwo
 \else \expandafter \@secondoftwo
 \fi
}%
\providecommand \@ifx [1]{%
 \ifx #1\expandafter \@firstoftwo
 \else \expandafter \@secondoftwo
 \fi
}%
\providecommand \natexlab [1]{#1}%
\providecommand \enquote  [1]{``#1''}%
\providecommand \bibnamefont  [1]{#1}%
\providecommand \bibfnamefont [1]{#1}%
\providecommand \citenamefont [1]{#1}%
\providecommand \href@noop [0]{\@secondoftwo}%
\providecommand \href [0]{\begingroup \@sanitize@url \@href}%
\providecommand \@href[1]{\@@startlink{#1}\@@href}%
\providecommand \@@href[1]{\endgroup#1\@@endlink}%
\providecommand \@sanitize@url [0]{\catcode `\\12\catcode `\$12\catcode `\&12\catcode `\#12\catcode `\^12\catcode `\_12\catcode `\%12\relax}%
\providecommand \@@startlink[1]{}%
\providecommand \@@endlink[0]{}%
\providecommand \url  [0]{\begingroup\@sanitize@url \@url }%
\providecommand \@url [1]{\endgroup\@href {#1}{\urlprefix }}%
\providecommand \urlprefix  [0]{URL }%
\providecommand \Eprint [0]{\href }%
\providecommand \doibase [0]{https://doi.org/}%
\providecommand \selectlanguage [0]{\@gobble}%
\providecommand \bibinfo  [0]{\@secondoftwo}%
\providecommand \bibfield  [0]{\@secondoftwo}%
\providecommand \translation [1]{[#1]}%
\providecommand \BibitemOpen [0]{}%
\providecommand \bibitemStop [0]{}%
\providecommand \bibitemNoStop [0]{.\EOS\space}%
\providecommand \EOS [0]{\spacefactor3000\relax}%
\providecommand \BibitemShut  [1]{\csname bibitem#1\endcsname}%
\let\auto@bib@innerbib\@empty
\bibitem [{\citenamefont {Niu}\ \emph {et~al.}(1985)\citenamefont {Niu}, \citenamefont {Thouless},\ and\ \citenamefont {Wu}}]{TKNN:1995}%
  \BibitemOpen
  \bibfield  {author} {\bibinfo {author} {\bibfnamefont {Q.}~\bibnamefont {Niu}}, \bibinfo {author} {\bibfnamefont {D.~J.}\ \bibnamefont {Thouless}},\ and\ \bibinfo {author} {\bibfnamefont {Y.-S.}\ \bibnamefont {Wu}},\ }\bibfield  {title} {\bibinfo {title} {Quantized hall conductance as a topological invariant},\ }\href {https://doi.org/10.1103/PhysRevB.31.3372} {\bibfield  {journal} {\bibinfo  {journal} {Phys. Rev. B}\ }\textbf {\bibinfo {volume} {31}},\ \bibinfo {pages} {3372} (\bibinfo {year} {1985})}\BibitemShut {NoStop}%
\bibitem [{\citenamefont {{Avron}}\ \emph {et~al.}(1995)\citenamefont {{Avron}}, \citenamefont {{Seiler}},\ and\ \citenamefont {{Zograf}}}]{avron:seiler:zograf:95}%
  \BibitemOpen
  \bibfield  {author} {\bibinfo {author} {\bibfnamefont {J.~E.}\ \bibnamefont {{Avron}}}, \bibinfo {author} {\bibfnamefont {R.}~\bibnamefont {{Seiler}}},\ and\ \bibinfo {author} {\bibfnamefont {P.~G.}\ \bibnamefont {{Zograf}}},\ }\bibfield  {title} {\bibinfo {title} {{Viscosity of Quantum Hall Fluids}},\ }\href {https://doi.org/10.1103/PhysRevLett.75.697} {\bibfield  {journal} {\bibinfo  {journal} {\prl}\ }\textbf {\bibinfo {volume} {75}},\ \bibinfo {pages} {697} (\bibinfo {year} {1995})},\ \Eprint {https://arxiv.org/abs/cond-mat/9502011} {arXiv:cond-mat/9502011 [cond-mat]} \BibitemShut {NoStop}%
\bibitem [{\citenamefont {Read}\ and\ \citenamefont {Rezayi}(2011)}]{Read:2011}%
  \BibitemOpen
  \bibfield  {author} {\bibinfo {author} {\bibfnamefont {N.}~\bibnamefont {Read}}\ and\ \bibinfo {author} {\bibfnamefont {E.~H.}\ \bibnamefont {Rezayi}},\ }\bibfield  {title} {\bibinfo {title} {Hall viscosity, orbital spin, and geometry: Paired superfluids and quantum hall systems},\ }\bibfield  {journal} {\bibinfo  {journal} {Physical Review B}\ }\textbf {\bibinfo {volume} {84}},\ \href {https://doi.org/10.1103/physrevb.84.085316} {10.1103/physrevb.84.085316} (\bibinfo {year} {2011})\BibitemShut {NoStop}%
\bibitem [{\citenamefont {Haldane}(2011)}]{Haldane:2011}%
  \BibitemOpen
  \bibfield  {author} {\bibinfo {author} {\bibfnamefont {F.~D.~M.}\ \bibnamefont {Haldane}},\ }\bibfield  {title} {\bibinfo {title} {Geometrical description of the fractional quantum hall effect},\ }\bibfield  {journal} {\bibinfo  {journal} {Physical Review Letters}\ }\textbf {\bibinfo {volume} {107}},\ \href {https://doi.org/10.1103/physrevlett.107.116801} {10.1103/physrevlett.107.116801} (\bibinfo {year} {2011})\BibitemShut {NoStop}%
\bibitem [{\citenamefont {Klevtsov}\ and\ \citenamefont {Wiegmann}(2015)}]{klevtsov:wiegmann:2015}%
  \BibitemOpen
  \bibfield  {author} {\bibinfo {author} {\bibfnamefont {S.}~\bibnamefont {Klevtsov}}\ and\ \bibinfo {author} {\bibfnamefont {P.}~\bibnamefont {Wiegmann}},\ }\bibfield  {title} {\bibinfo {title} {Geometric adiabatic transport in quantum hall states},\ }\href {https://doi.org/10.1103/PhysRevLett.115.086801} {\bibfield  {journal} {\bibinfo  {journal} {Phys. Rev. Lett.}\ }\textbf {\bibinfo {volume} {115}},\ \bibinfo {pages} {086801} (\bibinfo {year} {2015})}\BibitemShut {NoStop}%
\bibitem [{\citenamefont {Bradlyn}\ and\ \citenamefont {Read}(2015)}]{Bradlyn:2015}%
  \BibitemOpen
  \bibfield  {author} {\bibinfo {author} {\bibfnamefont {B.}~\bibnamefont {Bradlyn}}\ and\ \bibinfo {author} {\bibfnamefont {N.}~\bibnamefont {Read}},\ }\bibfield  {title} {\bibinfo {title} {Topological central charge from berry curvature: Gravitational anomalies in trial wave functions for topological phases},\ }\bibfield  {journal} {\bibinfo  {journal} {Physical Review B}\ }\textbf {\bibinfo {volume} {91}},\ \href {https://doi.org/10.1103/physrevb.91.165306} {10.1103/physrevb.91.165306} (\bibinfo {year} {2015})\BibitemShut {NoStop}%
\bibitem [{\citenamefont {Klevtsov}\ \emph {et~al.}(2016)\citenamefont {Klevtsov}, \citenamefont {Ma}, \citenamefont {Marinescu},\ and\ \citenamefont {Wiegmann}}]{Klevtsov:2016}%
  \BibitemOpen
  \bibfield  {author} {\bibinfo {author} {\bibfnamefont {S.}~\bibnamefont {Klevtsov}}, \bibinfo {author} {\bibfnamefont {X.}~\bibnamefont {Ma}}, \bibinfo {author} {\bibfnamefont {G.}~\bibnamefont {Marinescu}},\ and\ \bibinfo {author} {\bibfnamefont {P.}~\bibnamefont {Wiegmann}},\ }\bibfield  {title} {\bibinfo {title} {Quantum hall effect and quillen metric},\ }\href {https://doi.org/10.1007/s00220-016-2789-2} {\bibfield  {journal} {\bibinfo  {journal} {Communications in Mathematical Physics}\ }\textbf {\bibinfo {volume} {349}},\ \bibinfo {pages} {819–855} (\bibinfo {year} {2016})}\BibitemShut {NoStop}%
\bibitem [{\citenamefont {Liu}\ \emph {et~al.}(2024)\citenamefont {Liu}, \citenamefont {Mera}, \citenamefont {Fujimoto}, \citenamefont {Ozawa},\ and\ \citenamefont {Wang}}]{liu:mera:fujimoto:ozawa:wang:24}%
  \BibitemOpen
  \bibfield  {author} {\bibinfo {author} {\bibfnamefont {Z.}~\bibnamefont {Liu}}, \bibinfo {author} {\bibfnamefont {B.}~\bibnamefont {Mera}}, \bibinfo {author} {\bibfnamefont {M.}~\bibnamefont {Fujimoto}}, \bibinfo {author} {\bibfnamefont {T.}~\bibnamefont {Ozawa}},\ and\ \bibinfo {author} {\bibfnamefont {J.}~\bibnamefont {Wang}},\ }\href {https://arxiv.org/abs/2405.14479} {\bibinfo {title} {{Theory of Generalized Landau Levels and Implication for non-Abelian States}}} (\bibinfo {year} {2024}),\ \Eprint {https://arxiv.org/abs/2405.14479} {arXiv:2405.14479 [cond-mat.mes-hall]} \BibitemShut {NoStop}%
\bibitem [{\citenamefont {Fujimoto}\ \emph {et~al.}(2024)\citenamefont {Fujimoto}, \citenamefont {Parker}, \citenamefont {Dong}, \citenamefont {Khalaf}, \citenamefont {Vishwanath},\ and\ \citenamefont {Ledwith}}]{fujimoto2024higher}%
  \BibitemOpen
  \bibfield  {author} {\bibinfo {author} {\bibfnamefont {M.}~\bibnamefont {Fujimoto}}, \bibinfo {author} {\bibfnamefont {D.~E.}\ \bibnamefont {Parker}}, \bibinfo {author} {\bibfnamefont {J.}~\bibnamefont {Dong}}, \bibinfo {author} {\bibfnamefont {E.}~\bibnamefont {Khalaf}}, \bibinfo {author} {\bibfnamefont {A.}~\bibnamefont {Vishwanath}},\ and\ \bibinfo {author} {\bibfnamefont {P.}~\bibnamefont {Ledwith}},\ }\bibfield  {title} {\bibinfo {title} {Higher vortexability: zero field realization of higher landau levels},\ }\href {https://arxiv.org/abs/2403.00856} {\bibfield  {journal} {\bibinfo  {journal} {arXiv preprint arXiv:2403.00856}\ } (\bibinfo {year} {2024})}\BibitemShut {NoStop}%
\bibitem [{\citenamefont {Roy}(2014)}]{rahul:14}%
  \BibitemOpen
  \bibfield  {author} {\bibinfo {author} {\bibfnamefont {R.}~\bibnamefont {Roy}},\ }\bibfield  {title} {\bibinfo {title} {Band geometry of fractional topological insulators},\ }\href {https://doi.org/10.1103/PhysRevB.90.165139} {\bibfield  {journal} {\bibinfo  {journal} {Phys. Rev. B}\ }\textbf {\bibinfo {volume} {90}},\ \bibinfo {pages} {165139} (\bibinfo {year} {2014})}\BibitemShut {NoStop}%
\bibitem [{\citenamefont {Claassen}\ \emph {et~al.}(2015)\citenamefont {Claassen}, \citenamefont {Lee}, \citenamefont {Thomale}, \citenamefont {Qi},\ and\ \citenamefont {Devereaux}}]{claassen:15}%
  \BibitemOpen
  \bibfield  {author} {\bibinfo {author} {\bibfnamefont {M.}~\bibnamefont {Claassen}}, \bibinfo {author} {\bibfnamefont {C.~H.}\ \bibnamefont {Lee}}, \bibinfo {author} {\bibfnamefont {R.}~\bibnamefont {Thomale}}, \bibinfo {author} {\bibfnamefont {X.-L.}\ \bibnamefont {Qi}},\ and\ \bibinfo {author} {\bibfnamefont {T.~P.}\ \bibnamefont {Devereaux}},\ }\bibfield  {title} {\bibinfo {title} {Position-momentum duality and fractional quantum hall effect in chern insulators},\ }\href {https://doi.org/10.1103/PhysRevLett.114.236802} {\bibfield  {journal} {\bibinfo  {journal} {Phys. Rev. Lett.}\ }\textbf {\bibinfo {volume} {114}},\ \bibinfo {pages} {236802} (\bibinfo {year} {2015})}\BibitemShut {NoStop}%
\bibitem [{\citenamefont {Ozawa}\ and\ \citenamefont {Mera}(2021)}]{ozawa:mera:2021}%
  \BibitemOpen
  \bibfield  {author} {\bibinfo {author} {\bibfnamefont {T.}~\bibnamefont {Ozawa}}\ and\ \bibinfo {author} {\bibfnamefont {B.}~\bibnamefont {Mera}},\ }\bibfield  {title} {\bibinfo {title} {{Relations between topology and the quantum metric for Chern insulators}},\ }\href {https://doi.org/10.1103/PhysRevB.104.045103} {\bibfield  {journal} {\bibinfo  {journal} {Phys. Rev. B}\ }\textbf {\bibinfo {volume} {104}},\ \bibinfo {pages} {045103} (\bibinfo {year} {2021})}\BibitemShut {NoStop}%
\bibitem [{\citenamefont {Mera}\ and\ \citenamefont {Ozawa}(2021)}]{mera:ozawa:21}%
  \BibitemOpen
  \bibfield  {author} {\bibinfo {author} {\bibfnamefont {B.}~\bibnamefont {Mera}}\ and\ \bibinfo {author} {\bibfnamefont {T.}~\bibnamefont {Ozawa}},\ }\bibfield  {title} {\bibinfo {title} {{K\"ahler geometry and Chern insulators: Relations between topology and the quantum metric}},\ }\href {https://doi.org/10.1103/PhysRevB.104.045104} {\bibfield  {journal} {\bibinfo  {journal} {Phys. Rev. B}\ }\textbf {\bibinfo {volume} {104}},\ \bibinfo {pages} {045104} (\bibinfo {year} {2021})}\BibitemShut {NoStop}%
\bibitem [{\citenamefont {Wang}\ \emph {et~al.}(2021)\citenamefont {Wang}, \citenamefont {Cano}, \citenamefont {Millis}, \citenamefont {Liu},\ and\ \citenamefont {Yang}}]{jie:cano:millis:liu:yang:21}%
  \BibitemOpen
  \bibfield  {author} {\bibinfo {author} {\bibfnamefont {J.}~\bibnamefont {Wang}}, \bibinfo {author} {\bibfnamefont {J.}~\bibnamefont {Cano}}, \bibinfo {author} {\bibfnamefont {A.~J.}\ \bibnamefont {Millis}}, \bibinfo {author} {\bibfnamefont {Z.}~\bibnamefont {Liu}},\ and\ \bibinfo {author} {\bibfnamefont {B.}~\bibnamefont {Yang}},\ }\bibfield  {title} {\bibinfo {title} {Exact landau level description of geometry and interaction in a flatband},\ }\href {https://doi.org/10.1103/PhysRevLett.127.246403} {\bibfield  {journal} {\bibinfo  {journal} {Phys. Rev. Lett.}\ }\textbf {\bibinfo {volume} {127}},\ \bibinfo {pages} {246403} (\bibinfo {year} {2021})}\BibitemShut {NoStop}%
\bibitem [{\citenamefont {Ledwith}\ \emph {et~al.}(2020)\citenamefont {Ledwith}, \citenamefont {Tarnopolsky}, \citenamefont {Khalaf},\ and\ \citenamefont {Vishwanath}}]{ledwith:tarnopolsky:khalaf:vishwanath:20}%
  \BibitemOpen
  \bibfield  {author} {\bibinfo {author} {\bibfnamefont {P.~J.}\ \bibnamefont {Ledwith}}, \bibinfo {author} {\bibfnamefont {G.}~\bibnamefont {Tarnopolsky}}, \bibinfo {author} {\bibfnamefont {E.}~\bibnamefont {Khalaf}},\ and\ \bibinfo {author} {\bibfnamefont {A.}~\bibnamefont {Vishwanath}},\ }\bibfield  {title} {\bibinfo {title} {Fractional chern insulator states in twisted bilayer graphene: An analytical approach},\ }\href {https://doi.org/10.1103/PhysRevResearch.2.023237} {\bibfield  {journal} {\bibinfo  {journal} {Phys. Rev. Research}\ }\textbf {\bibinfo {volume} {2}},\ \bibinfo {pages} {023237} (\bibinfo {year} {2020})}\BibitemShut {NoStop}%
\bibitem [{\citenamefont {{Ledwith}}\ \emph {et~al.}(2022)\citenamefont {{Ledwith}}, \citenamefont {{Vishwanath}},\ and\ \citenamefont {{Parker}}}]{ledwith:vishwanath:parker:22}%
  \BibitemOpen
  \bibfield  {author} {\bibinfo {author} {\bibfnamefont {P.~J.}\ \bibnamefont {{Ledwith}}}, \bibinfo {author} {\bibfnamefont {A.}~\bibnamefont {{Vishwanath}}},\ and\ \bibinfo {author} {\bibfnamefont {D.~E.}\ \bibnamefont {{Parker}}},\ }\bibfield  {title} {\bibinfo {title} {{Vortexability: A Unifying Criterion for Ideal Fractional Chern Insulators}},\ }\href@noop {} {\bibfield  {journal} {\bibinfo  {journal} {arXiv e-prints}\ ,\ \bibinfo {eid} {arXiv:2209.15023}} (\bibinfo {year} {2022})},\ \Eprint {https://arxiv.org/abs/2209.15023} {arXiv:2209.15023 [cond-mat.str-el]} \BibitemShut {NoStop}%
\bibitem [{\citenamefont {Wang}\ and\ \citenamefont {Liu}(2022)}]{jie:liu:22}%
  \BibitemOpen
  \bibfield  {author} {\bibinfo {author} {\bibfnamefont {J.}~\bibnamefont {Wang}}\ and\ \bibinfo {author} {\bibfnamefont {Z.}~\bibnamefont {Liu}},\ }\bibfield  {title} {\bibinfo {title} {Hierarchy of ideal flatbands in chiral twisted multilayer graphene models},\ }\href {https://doi.org/10.1103/PhysRevLett.128.176403} {\bibfield  {journal} {\bibinfo  {journal} {Phys. Rev. Lett.}\ }\textbf {\bibinfo {volume} {128}},\ \bibinfo {pages} {176403} (\bibinfo {year} {2022})}\BibitemShut {NoStop}%
\bibitem [{\citenamefont {Ledwith}\ \emph {et~al.}(2022)\citenamefont {Ledwith}, \citenamefont {Vishwanath},\ and\ \citenamefont {Khalaf}}]{ledwith:vishwanath:khalaf:22}%
  \BibitemOpen
  \bibfield  {author} {\bibinfo {author} {\bibfnamefont {P.~J.}\ \bibnamefont {Ledwith}}, \bibinfo {author} {\bibfnamefont {A.}~\bibnamefont {Vishwanath}},\ and\ \bibinfo {author} {\bibfnamefont {E.}~\bibnamefont {Khalaf}},\ }\bibfield  {title} {\bibinfo {title} {Family of ideal chern flatbands with arbitrary chern number in chiral twisted graphene multilayers},\ }\href {https://doi.org/10.1103/PhysRevLett.128.176404} {\bibfield  {journal} {\bibinfo  {journal} {Phys. Rev. Lett.}\ }\textbf {\bibinfo {volume} {128}},\ \bibinfo {pages} {176404} (\bibinfo {year} {2022})}\BibitemShut {NoStop}%
\bibitem [{\citenamefont {{Wang}}\ \emph {et~al.}(2024)\citenamefont {{Wang}}, \citenamefont {{Zhang}}, \citenamefont {{Liu}}, \citenamefont {{Wang}}, \citenamefont {{Cao}},\ and\ \citenamefont {{Xiao}}}]{nonabelian:di:24}%
  \BibitemOpen
  \bibfield  {author} {\bibinfo {author} {\bibfnamefont {C.}~\bibnamefont {{Wang}}}, \bibinfo {author} {\bibfnamefont {X.-W.}\ \bibnamefont {{Zhang}}}, \bibinfo {author} {\bibfnamefont {X.}~\bibnamefont {{Liu}}}, \bibinfo {author} {\bibfnamefont {J.}~\bibnamefont {{Wang}}}, \bibinfo {author} {\bibfnamefont {T.}~\bibnamefont {{Cao}}},\ and\ \bibinfo {author} {\bibfnamefont {D.}~\bibnamefont {{Xiao}}},\ }\bibfield  {title} {\bibinfo {title} {{Higher Landau-Level Analogues and Signatures of Non-Abelian States in Twisted Bilayer MoTe$_2$}},\ }\href {https://doi.org/10.48550/arXiv.2404.05697} {\bibfield  {journal} {\bibinfo  {journal} {arXiv e-prints}\ ,\ \bibinfo {eid} {arXiv:2404.05697}} (\bibinfo {year} {2024})},\ \Eprint {https://arxiv.org/abs/2404.05697} {arXiv:2404.05697 [cond-mat.str-el]} \BibitemShut {NoStop}%
\bibitem [{\citenamefont {Ahn}\ \emph {et~al.}(2024)\citenamefont {Ahn}, \citenamefont {Lee}, \citenamefont {Yananose}, \citenamefont {Kim},\ and\ \citenamefont {Cho}}]{nonabelian:cho:24}%
  \BibitemOpen
  \bibfield  {author} {\bibinfo {author} {\bibfnamefont {C.-E.}\ \bibnamefont {Ahn}}, \bibinfo {author} {\bibfnamefont {W.}~\bibnamefont {Lee}}, \bibinfo {author} {\bibfnamefont {K.}~\bibnamefont {Yananose}}, \bibinfo {author} {\bibfnamefont {Y.}~\bibnamefont {Kim}},\ and\ \bibinfo {author} {\bibfnamefont {G.~Y.}\ \bibnamefont {Cho}},\ }\bibfield  {title} {\bibinfo {title} {Non-abelian fractional quantum anomalous hall states and first landau level physics of the second moir\'e band of twisted bilayer ${\mathrm{mote}}_{2}$},\ }\href {https://doi.org/10.1103/PhysRevB.110.L161109} {\bibfield  {journal} {\bibinfo  {journal} {Phys. Rev. B}\ }\textbf {\bibinfo {volume} {110}},\ \bibinfo {pages} {L161109} (\bibinfo {year} {2024})}\BibitemShut {NoStop}%
\bibitem [{\citenamefont {{Xu}}\ \emph {et~al.}(2024)\citenamefont {{Xu}}, \citenamefont {{Mao}}, \citenamefont {{Zeng}},\ and\ \citenamefont {{Zhang}}}]{nonabelian:zhang:24}%
  \BibitemOpen
  \bibfield  {author} {\bibinfo {author} {\bibfnamefont {C.}~\bibnamefont {{Xu}}}, \bibinfo {author} {\bibfnamefont {N.}~\bibnamefont {{Mao}}}, \bibinfo {author} {\bibfnamefont {T.}~\bibnamefont {{Zeng}}},\ and\ \bibinfo {author} {\bibfnamefont {Y.}~\bibnamefont {{Zhang}}},\ }\bibfield  {title} {\bibinfo {title} {{Multiple Chern bands in twisted MoTe$_2$ and possible non-Abelian states}},\ }\href {https://doi.org/10.48550/arXiv.2403.17003} {\bibfield  {journal} {\bibinfo  {journal} {arXiv e-prints}\ ,\ \bibinfo {eid} {arXiv:2403.17003}} (\bibinfo {year} {2024})},\ \Eprint {https://arxiv.org/abs/2403.17003} {arXiv:2403.17003 [cond-mat.str-el]} \BibitemShut {NoStop}%
\bibitem [{\citenamefont {Reddy}\ \emph {et~al.}(2024)\citenamefont {Reddy}, \citenamefont {Paul}, \citenamefont {Abouelkomsan},\ and\ \citenamefont {Fu}}]{nonabelian:fu:24}%
  \BibitemOpen
  \bibfield  {author} {\bibinfo {author} {\bibfnamefont {A.~P.}\ \bibnamefont {Reddy}}, \bibinfo {author} {\bibfnamefont {N.}~\bibnamefont {Paul}}, \bibinfo {author} {\bibfnamefont {A.}~\bibnamefont {Abouelkomsan}},\ and\ \bibinfo {author} {\bibfnamefont {L.}~\bibnamefont {Fu}},\ }\bibfield  {title} {\bibinfo {title} {Non-abelian fractionalization in topological minibands},\ }\href {https://doi.org/10.1103/PhysRevLett.133.166503} {\bibfield  {journal} {\bibinfo  {journal} {Phys. Rev. Lett.}\ }\textbf {\bibinfo {volume} {133}},\ \bibinfo {pages} {166503} (\bibinfo {year} {2024})}\BibitemShut {NoStop}%
\bibitem [{\citenamefont {Onishi}\ and\ \citenamefont {Fu}(2024)}]{onishi:fu:24}%
  \BibitemOpen
  \bibfield  {author} {\bibinfo {author} {\bibfnamefont {Y.}~\bibnamefont {Onishi}}\ and\ \bibinfo {author} {\bibfnamefont {L.}~\bibnamefont {Fu}},\ }\bibfield  {title} {\bibinfo {title} {{Topological Bound on the Structure Factor}},\ }\href {https://doi.org/10.1103/PhysRevLett.133.206602} {\bibfield  {journal} {\bibinfo  {journal} {Phys. Rev. Lett.}\ }\textbf {\bibinfo {volume} {133}},\ \bibinfo {pages} {206602} (\bibinfo {year} {2024})}\BibitemShut {NoStop}%
\bibitem [{\citenamefont {Mera}(2020)}]{mera:2020}%
  \BibitemOpen
  \bibfield  {author} {\bibinfo {author} {\bibfnamefont {B.}~\bibnamefont {Mera}},\ }\bibfield  {title} {\bibinfo {title} {{Localization anisotropy and complex geometry in two-dimensional insulators}},\ }\href {https://doi.org/10.1103/PhysRevB.101.115128} {\bibfield  {journal} {\bibinfo  {journal} {Phys. Rev. B}\ }\textbf {\bibinfo {volume} {101}},\ \bibinfo {pages} {115128} (\bibinfo {year} {2020})}\BibitemShut {NoStop}%
\bibitem [{\citenamefont {Eells}\ and\ \citenamefont {Wood}(1983)}]{eells:wood:83}%
  \BibitemOpen
  \bibfield  {author} {\bibinfo {author} {\bibfnamefont {J.}~\bibnamefont {Eells}}\ and\ \bibinfo {author} {\bibfnamefont {C.}~\bibnamefont {Wood}},\ }\bibfield  {title} {\bibinfo {title} {{Harmonic maps from surfaces to complex projective spaces}},\ }\href {https://doi.org/https://doi.org/10.1016/0001-8708(83)90062-2} {\bibfield  {journal} {\bibinfo  {journal} {Advances in Mathematics}\ }\textbf {\bibinfo {volume} {49}},\ \bibinfo {pages} {217} (\bibinfo {year} {1983})}\BibitemShut {NoStop}%
\bibitem [{Note1()}]{Note1}%
  \BibitemOpen
  \bibinfo {note} {The band is such that the image of the associated map to projective space is not contained in any proper linear subspace. This condition can always be met by restricting the total number of bands in the system.}\BibitemShut {Stop}%
\bibitem [{\citenamefont {Onishi}\ \emph {et~al.}(2024)\citenamefont {Onishi}, \citenamefont {Avdoshkin},\ and\ \citenamefont {Fu}}]{onishi:avdoshkin:fu:2024}%
  \BibitemOpen
  \bibfield  {author} {\bibinfo {author} {\bibfnamefont {Y.}~\bibnamefont {Onishi}}, \bibinfo {author} {\bibfnamefont {A.}~\bibnamefont {Avdoshkin}},\ and\ \bibinfo {author} {\bibfnamefont {L.}~\bibnamefont {Fu}},\ }\href {https://arxiv.org/abs/2412.02656} {\bibinfo {title} {{Geometric bound on structure factor}}} (\bibinfo {year} {2024}),\ \Eprint {https://arxiv.org/abs/2412.02656} {arXiv:2412.02656 [cond-mat.mes-hall]} \BibitemShut {NoStop}%
\bibitem [{Note2()}]{Note2}%
  \BibitemOpen
  \bibinfo {note} {The vanishing of the sum in the integrand, due to orthogonality of $P$ and $Q$, implies the vanishing of each of them and, also, they are the Hermitian conjugate of each other.}\BibitemShut {Stop}%
\bibitem [{\citenamefont {Calabi}(1953)}]{calabi:53}%
  \BibitemOpen
  \bibfield  {author} {\bibinfo {author} {\bibfnamefont {E.}~\bibnamefont {Calabi}},\ }\bibfield  {title} {\bibinfo {title} {Isometric imbedding of complex manifolds},\ }\href {http://www.jstor.org/stable/1969817} {\bibfield  {journal} {\bibinfo  {journal} {Annals of Mathematics}\ }\textbf {\bibinfo {volume} {58}},\ \bibinfo {pages} {1} (\bibinfo {year} {1953})}\BibitemShut {NoStop}%
\bibitem [{\citenamefont {Klevtsov}(2016)}]{klevtsov:lecturenotes:2016}%
  \BibitemOpen
  \bibfield  {author} {\bibinfo {author} {\bibfnamefont {S.}~\bibnamefont {Klevtsov}},\ }\bibfield  {title} {\bibinfo {title} {{Geometry and large N limits in Laughlin states}},\ }\href@noop {} {\bibfield  {journal} {\bibinfo  {journal} {Travaux Mathematiques}\ }\textbf {\bibinfo {volume} {24}},\ \bibinfo {pages} {63} (\bibinfo {year} {2016})}\BibitemShut {NoStop}%
\bibitem [{\citenamefont {Lawson}(1980)}]{lawson:1980}%
  \BibitemOpen
  \bibfield  {author} {\bibinfo {author} {\bibfnamefont {H.~B.}\ \bibnamefont {Lawson}},\ }\href@noop {} {\emph {\bibinfo {title} {{Lectures on minimal submanifolds}}}}\ (\bibinfo  {publisher} {Inst. for Pure-Appl. Math.},\ \bibinfo {year} {1980})\BibitemShut {NoStop}%
\bibitem [{\citenamefont {Green}(1978)}]{green:1978}%
  \BibitemOpen
  \bibfield  {author} {\bibinfo {author} {\bibfnamefont {M.~L.}\ \bibnamefont {Green}},\ }\bibfield  {title} {\bibinfo {title} {{Metric rigidity of holomorphic maps to K{\"a}hler manifolds}},\ }\href@noop {} {\bibfield  {journal} {\bibinfo  {journal} {Journal of Differential Geometry}\ }\textbf {\bibinfo {volume} {13}},\ \bibinfo {pages} {279} (\bibinfo {year} {1978})}\BibitemShut {NoStop}%
\end{thebibliography}%

\appendix
\begin{widetext}
\section*{--- Appendix ---}
\section{Proof that GLLs are Harmonic maps}
We start by showing that ideal K\"ahler bands are harmonic. Namely, the holomorphicity condition of the K\"ahler band, as established in Ref.~\cite{mera:ozawa:21}, implies harmonicity:
\begin{align}
Q\frac{\partial P}{\partial \bar{z}} P= 0\implies Q\frac{\partial^2 P}{\partial z \partial \bar{z}}P=0.
\end{align}
To verify this, differentiate the first expression with respect to $z$:
\begin{align}
0 &=\frac{\partial}{\partial z}\left(Q\frac{\partial P}{\partial \bar{z}} P\right)\nonumber \\
&=-\frac{\partial P}{\partial z}\frac{\partial P}{\partial \bar{z}}P +Q\frac{\partial^2 P}{\partial z\partial \bar{z}} P +Q \frac{\partial P}{\partial \bar{z}}\frac{\partial P}{\partial z}=Q\frac{\partial^2 P}{\partial z\partial \bar{z}} P,
\end{align}
where we used $QdP=dPP=QdPP$ and the holomorphicity condition. 

Next, we show that all the $P_i$'s are all critical points of $E$ and hence determine harmonic maps. First, choosing a unit-norm vector $\ket{u}$ representing $P$, i.e., $P=\ket{u}\bra{u}$, we show that Eq.~\eqref{eq: harmonic band projector equation} is equivalent to
\begin{align}
Q\left(\frac{\partial^2}{\partial z \partial \bar{z}} \ket{u} -\bra{u}\frac{\partial}{\partial z}\ket{u} \frac{\partial}{\partial \bar{z}}\ket{u}-\bra{u}\frac{\partial}{\partial \bar{z}}\ket{u} \frac{\partial}{\partial z}\ket{u}\right)=0.
\label{eq: harmonic band wavefunction equation}
\end{align}
Explicitly
\begin{align}
Q\frac{\partial^2 P}{\partial z \partial \bar{z}}P &=Q\frac{\partial}{\partial z}\left(\frac{\partial \ket{u}}{\partial \bar{z}}\bra{u}+\ket{u}\frac{\partial \bra{u}}{\partial \bar{z}}\right)P \nonumber\\
&=Q\left(\frac{\partial^2 \ket{u}}{\partial z \partial \bar{z}}\bra{u} +\frac{\partial \ket{u}}{\partial \bar{z}}\frac{\partial \bra{u}}{\partial z}+\frac{\partial \ket{u}}{\partial z}\frac{\partial \bra{u}}{\partial \bar{z}}\right)P \nonumber\\
&=Q\left(\frac{\partial^2 \ket{u}}{\partial z \partial \bar{z}} -\bra{u}\frac{\partial}{\partial z}\ket{u}\frac{\partial \ket{u}}{\partial \bar{z}}-\bra{u}\frac{\partial }{\partial \bar{z}}\ket{u}\frac{\partial \ket{u}}{\partial z}\right)\bra{u},
\end{align}
where we used $Q\ket{u}=0$ and the fact that $\ket{u}$ has unit norm so that $\bra{u}d\ket{u}=-d\left(\bra{u}\right)\ket{u}$. Finally, taking the inner product with $\ket{u}$, we get the desired result. Next, we set $\ket{u}=\ket{u_i}$ and recall the Frenet-Serret Equations, Eq.~\eqref{eq: Frenet-Serret equations}, which imply, taking into account that $\theta^{i+1}_{\; i}$ is a $(1,0)$-form and $\theta^{i-1}_{\; i}$ a $(0,1)$-form,
\begin{align}
\frac{\partial \ket{u_i}}{\partial z} &= \ket{u_i}\theta^{i}_{\; i}\left(\frac{\partial}{\partial z}\right) +\ket{u_{i+1}}\theta^{i+1}_{\; i}\left(\frac{\partial}{\partial z}\right).
\label{eq: holomorphic derivative of ui}
\end{align}
Similarly,
\begin{align}
\frac{\partial \ket{u_i}}{\partial \bar{z}} &= \ket{u_{i-1}}\theta^{i-1}_{\; i}\left(\frac{\partial}{\partial \bar{z}}\right) +\ket{u_i}\theta^{i}_{\; i}\left(\frac{\partial}{\partial \bar{z}}\right),
\label{eq: antiholomorphic derivative of ui}
\end{align}
where the notation $\alpha(X)=a_ib^i$ denotes the evaluation of the one-form $\alpha=a_idk^i$ in a vector field $X=b^i\frac{\partial}{\partial k^i}$. Taking the derivative of Eq.~\eqref{eq: holomorphic derivative of ui} with respect to $\bar{z}$ and using Eq.~\eqref{eq: antiholomorphic derivative of ui}, we can get
\begin{align}
Q\frac{\partial^2\ket{u_i}}{ \partial \bar{z} \partial z } &= Q\left(\ket{u_{i-1}}\theta^{i-1}_{\; i}\left(\frac{\partial}{\partial \bar{z}}\right) +\ket{u_i}\theta^{i}_{\; i}\left(\frac{\partial}{\partial \bar{z}}\right)\right)\theta^{i}_{\; i}\left(\frac{\partial}{\partial z}\right) \nonumber \\
&+ Q\ket{u_{i}}\frac{\partial}{\partial \bar{z}}\left(\theta^{i-1}_{\; i}\left(\frac{\partial}{\partial \bar{z}}\right)\right) \nonumber \\
&+Q\left(\ket{u_{i}}\theta^{i}_{\; i+1}\left(\frac{\partial}{\partial \bar{z}}\right) +\ket{u_{i+1}}\theta^{i+1}_{\; i+1}\left(\frac{\partial}{\partial \bar{z}}\right)\right)\theta^{i+1}_{\;i}\left(\frac{\partial}{\partial z}\right) \nonumber \\
&+Q\ket{u_{i+1}}\frac{\partial}{\partial\bar{z}}\left(\theta^{i+1}_{\;i}\left(\frac{\partial}{\partial z}\right)\right) \nonumber \\
&=\ket{u_{i-1}}\theta^{i-1}_{\; i}\left(\frac{\partial}{\partial \bar{z}}\right)\theta^{i}_{\; i}\left(\frac{\partial}{\partial z}\right)\nonumber \\
&+\ket{u_{i+1}}\left(\theta^{i+1}_{\; i+1}\left(\frac{\partial}{\partial \bar{z}}\right)\theta^{i+1}_{\;i}\left(\frac{\partial}{\partial z}\right)+\frac{\partial}{\partial\bar{z}}\left(\theta^{i+1}_{\;i}\left(\frac{\partial}{\partial z}\right)\right)\right)
\end{align}
where we used $Q=\sum_{j\neq i}\ket{u_j}\bra{u_j}$. Note that because of smoothness of $\ket{u}$ this is the same as $Q\frac{\partial^2 \ket{u_i}}{\partial z\partial \bar{z}}$. The other two terms in the left-hand side of Eq.~\eqref{eq: harmonic band wavefunction equation} are, using again Eq.~\eqref{eq: holomorphic derivative of ui} and Eq.~\eqref{eq: antiholomorphic derivative of ui} and $Q=\sum_{j\neq i} \ket{u_j}\bra{u_j}$,
\begin{align}
-\bra{u_i}\frac{\partial}{\partial z}\ket{u_i}Q\frac{\partial \ket{u_i}}{\partial \bar{z}} &=-\theta^{i}_{\; i}\left(\frac{\partial}{\partial z} \right) Q\left(\ket{u_{i-1}}\theta^{i-1}_{\; i}\left(\frac{\partial}{\partial \bar{z}}\right) +\ket{u_i}\theta^{i}_{\; i}\left(\frac{\partial}{\partial \bar{z}}\right)\right) \nonumber \\
&=-\ket{u_{i-1}}\theta^{i}_{\; i}\left(\frac{\partial}{\partial z} \right) \theta^{i-1}_{\; i}\left(\frac{\partial}{\partial \bar{z}}\right),
\end{align}
and, similarly,
\begin{align}
-\bra{u_i}\frac{\partial}{\partial \bar{z}}\ket{u_i}Q\frac{\partial \ket{u_i}}{\partial z} &=-\theta^{i}_{\; i}\left(\frac{\partial}{\partial \bar{z}}\right) Q\left(\ket{u_i}\theta^{i}_{\; i}\left(\frac{\partial}{\partial z}\right) +\ket{u_{i+1}}\theta^{i+1}_{\; i}\left(\frac{\partial}{\partial z}\right)\right) \nonumber \\
&=-\ket{u_{i+1}}\theta^{i}_{\; i}\left(\frac{\partial}{\partial \bar{z}}\right) \theta^{i+1}_{\; i}\left(\frac{\partial}{\partial z}\right).
\end{align}
Hence, plugging the above results in the left-hand side of Eq.~\eqref{eq: harmonic band wavefunction equation}, we find
\begin{align}
&Q\left(\frac{\partial^2}{\partial z \partial \bar{z}} \ket{u_i} -\bra{u_i}\frac{\partial}{\partial z}\ket{u_i} \frac{\partial}{\partial \bar{z}}\ket{u_i}-\bra{u_i}\frac{\partial}{\partial \bar{z}}\ket{u_i} \frac{\partial}{\partial z}\ket{u_i}\right)\nonumber\\
&=\ket{u_{i+1}}\Big[\frac{\partial}{\partial\bar{z}}\left(\theta^{i+1}_{\;i}\left(\frac{\partial}{\partial z}\right)\right)+\left(\theta^{i+1}_{\; i+1}\left(\frac{\partial}{\partial \bar{z}}\right)-\theta^{i}_{\; i}\left(\frac{\partial}{\partial \bar{z}}\right)\right)\theta^{i+1}_{\;i}\left(\frac{\partial}{\partial z}\right)\Big].
\end{align}
Finally, we claim that the factor in parenthesis satisfies
\begin{align}
\frac{\partial}{\partial\bar{z}}\left(\theta^{i+1}_{\;i}\left(\frac{\partial}{\partial z}\right)\right)+\left(\theta^{i+1}_{\; i+1}\left(\frac{\partial}{\partial \bar{z}}\right)-\theta^{i}_{\; i}\left(\frac{\partial}{\partial \bar{z}}\right)\right)\theta^{i+1}_{\;i}\left(\frac{\partial}{\partial z}\right)=0, 
\label{eq: factor that cancels proving harmonicity}
\end{align}
which implies the result. To prove this, we use the Cartan structure equation
\begin{align}
d\theta^{i+1}_{\; i} &=-\sum_{j} \theta^{i+1}_{\; j}\wedge \theta^{j}_{\; i}=-\theta^{i+1}_{\; i}\wedge \theta^{i}_{\; i} -\theta^{i+1}_{\; i+1}\wedge \theta^{i+1}_{\; i}\nonumber \\
&=  \left(\theta^{i}_{\; i} -\theta^{i+1}_{\; i+1}\right)\wedge \theta^{i+1}_{\; i},
\end{align}
and contract it with $\frac{\partial}{\partial \bar{z}}$ and $\frac{\partial}{\partial z}$. The left-hand side of the equation gives
\begin{align}
d\theta^{i+1}_{\; i}\left(\frac{\partial}{\partial \bar{z}},\frac{\partial}{\partial z}\right)=\frac{\partial}{\partial\bar{z}}\left(\theta^{i+1}_{\;i}\left(\frac{\partial}{\partial z}\right)\right)
\end{align}
and the right-hand side gives
\begin{align}
&\left[\left(\theta^{i}_{\; i} -\theta^{i+1}_{\; i+1}\right)\wedge \theta^{i+1}_{\; i}\right]\left(\frac{\partial}{\partial \bar{z}},\frac{\partial}{\partial z}\right)=\left(\theta^{i}_{\; i} \left(\frac{\partial}{\partial\bar{z}}\right)-\theta^{i+1}_{\; i+1}\left(\frac{\partial}{\partial \bar{z}}\right)\right)\theta^{i+1}_{\;i}\left(\frac{\partial}{\partial z}\right),
\end{align}
where we used the fact that $\theta^{i+1}_{\; i}$ is a $(1,0)$-form. Putting the two together implies Eq.~\eqref{eq: factor that cancels proving harmonicity}, which implies the desired result.

As described in the main text, the results of Eells and Wood from Ref.~\cite{eells:wood:83} allow us to go beyond what is presented above. Their Theorem~6.9. together with their Proposition~7.6. imply---assuming a fullness condition in that the band is such that the image of the associated map to projective space is not contained in any proper linear subspace [can always do that by restricting the total number of bands in the system]---that any harmonic maps with $\mathcal{C}\neq 0$ is described by a Bloch wavefunction $u_i$, for some $i$, appearing in the distinguished unitary moving frame of an ideal K\"ahler band. Here we should actually refer to the $u_i$'s as ideal GLLs in the same way that we refer to $u_0$ as an ideal K\"ahler band. This is because just as the ideal K\"ahler band is ideal in the sense that there is a fixed complex structure $J$ which is translation invariant, the $u_i$'s are ideal because there is a choice of a translation-invariant $h$ (for which $J$ is the natural complex structure built out of rotation by 90 degrees in positive orientation) defined up to scale. All other GLLs are obtained by pullback by an appropriate diffeomorphism. In this operation the complex structure and the metric are no longer translation-invariant.
\section{Proof of Calabi's rigidity theorem for the K\"ahler band setting}
\label{sec:}
Recall that a K\"ahler band is a holomorphic immersion to projective space described by a nonvanishing holomorphic vector $\ket{u_{\bf{k}}}=(f_1(z),\dots, f_N(z)) \in \mathbb{C}^N$---the Bloch wavefunction---, where the $f_i$'s are holomorphic functions with respect to some complex structure in the Brillouin zone. For simplicity, we may assume this complex structure associated to the complex coordinate $z=-k^y+\tau k^x$, $\tau\in\mathbb{H}$, as in the main text, but the result is independent of this choice. Under suitable boundary conditions for $\ket{u_{\bf{k}}}$, we can assume it to be a global function (it arises as global section of the pullback of the Bloch bundle by the quotient map $p:\mathbb{R}^2\to \BZ^2$, and hence it is a multi-valued global section of the Bloch bundle). Now Calabi's rigidity theorem in the K\"ahler band setting is as follows. If $\ket{u_{\bf{k}}}\in\mathbb{C}^N$ and $\ket{u'_{\bf{k}}}\in\mathbb{C}^{N+M}$, where $M$ is an arbitrary nonnegative integer, are K\"ahler bands determining the same quantum metric, then, there exists a unitary operator $U\in\mathrm{U}(N+M)$ such that, up to multiplication by a global nonvanishing holomorphic function, $\ket{u'_{\bf{k}}}=U\ket{u_{\bf{k}}}$, where we use the standard inclusion $\mathbb{C}^N\subset \mathbb{C}^{N+M}$, where a vector $(v_1,\dots,v_N)\in\mathbb{C}^N$ is mapped to $(v_1,\dots, v_N,0,\dots, 0)\in \mathbb{C}^{N+M}$. In other words, the induced holomorphic maps differ only by an isometry in projective space as a K\"ahler manifold. To prove this note that the equality of the quantum metrics, due to holomorphicity, is equivalent to the statement
\begin{align}
\partial \overline{\partial} \log \left(\langle u_{\bf{k}}|u_{\bf{k}}\rangle\right)-\partial\overline{\partial} \log \left(\langle u'_{\bf{k}}|u'_{\bf{k}}\rangle\right) =0 \iff \frac{\partial^2 \log \left(\frac{\langle u_{\bf{k}}|u_{\bf{k}}\rangle}{\langle u'_{\bf{k}}|u'_{\bf{k}}\rangle}\right)}{\partial z \partial \bar z} =0,
\end{align}
which implies that $\log \left(\frac{\langle u_{\bf{k}}|u_{\bf{k}}\rangle}{\langle u'_{\bf{k}}|u'_{\bf{k}}\rangle}\right)$ is the real part of some holomorphic function, i.e.,
\begin{align}
\langle u_{\bf{k}}|u_{\bf{k}}\rangle=\langle u'_{\bf{k}}|u'_{\bf{k}}\rangle e^{\alpha(z) +\overline{\alpha(z)}},
\end{align}
some holomorphic fuction $\alpha(z)$. We can multiply $\ket{u'_{\bf{k}}}$ by $e^{\alpha(z)}$, which is a nonvanishing holomorphic function, and this still determines the same band. Without loss of generality what we need to show is then that
if \begin{align}
\langle u_{\bf{k}}|u_{\bf{k}}\rangle=\langle u'_{\bf{k}}|u'_{\bf{k}}\rangle
\end{align}
holds for all momenta, then it implies the existence of a constant unitary relating them. Write $\ket{u_{\bf{k}}}=(f_1(z),\dots, f_N(z))=:f(z)$ and $\ket{u'_{\bf{k}}}=(g_1(z),\dots,g_{N+M}(z))=:g(z)$, then the above is equivalent to 
\begin{align}
\sum_{i=1}^{N}|f_i(z)|^2=\sum_{i=1}^{N+M}|g_i(z)|^2.\end{align}
We will prove the identity by induction on $N$. For $N=1$, since $f_1(z)\neq 0$, we may divide $(g_1(z),\dots,g_{1+M}(z))$ by $f_1(z)$ to define a new vector $\Phi(z)=g(z)/f_1(z)$, that satisfies
\begin{align*}
1=\sum_{i=1}^{N+1}|\Phi_i(z)|^2    
\end{align*}
Acting with $\partial^2/\partial z \partial \bar{z}$, we get the relation
\begin{align}
0=\sum_{i=1}^{N+1}|\Phi'_i(z)|^2 
\end{align}
This implies that $\Phi'(z)=0$, i.e., that $\Phi$ is a constant unit vector $b\in\mathbb{C}^{1+M}$ and thus, $g(z)=f_1(z)b$. But since $b$ is a unit vector, there exists a unitary $U\in\mathrm{U}(1+M)$ such that $g(z)=U(f_1(z),0,\dots,0)$. Hence the result is proved for $N=1$. Now let us assume the result holds up to $N-1$ and from that show it is valid for $N$. By rotating the basis of $\mathbb{C}^{N}$ we may assume that $(f_1(z),\dots, f_N(z))$ is such that $f_1(0)\neq 0$ and $f_2'(0)=\dots= f_{N}'(0)=0$. The relation $f_1(0)\neq 0$ holds in a small neighborhood of the origin due to continuity. Define $\Phi(z)=g(z)/f_1(z)\in\mathbb{C}^{N+M}$ and $\Psi(z)=(f_2(z)/f_1(z),\dots, f_N(z)/f_1(z))\in \mathbb{C}^{N-1}$. In that neighborhood we have the equation
\begin{align}
1 + \sum_{i=1}^{N-1}|\Psi_i(z)|^2 =   \sum_{i=1}^{N+M}|\Phi_i(z)|^2. 
\label{eq: Psi and Phi norms squared}
\end{align}
Acting with $\partial^2/\partial z \partial \bar{z}$, we get the relation
\begin{align}
\sum_{i=1}^{N-1}|\Psi'_i(z)|^2 =\sum_{i=1}^{N+M}|\Phi'_i(z)|^2. 
\end{align}
By the induction step, there exists a unitary $U\in\mathrm{U}(N+M)$ such that $\Phi'(z)=U(\Psi'_1(z),\dots, \Psi'_{N-1}(z),0,\dots,0)=:U\Psi'(z)$. Integrating this equation yields
\begin{align}
\Phi(z)=U\Psi(z)+b,  
\end{align}
for some constant vector $b$. Now we square the equation to obtain
\begin{align}
||\Phi(z)||^2= ||\Psi(z)||^2 + 2\mathrm{Re}\langle b|\Psi(z)\rangle + ||b||^2,   
\end{align}
but by Eq.~\eqref{eq: Psi and Phi norms squared}, we conclude that
\begin{align}
2\mathrm{Re}\langle b|\Psi(z)\rangle + ||b||^2=1.  
\end{align}
Observe that evaluating this at $z=0$, since $\Psi(0)=0$, we have $||b||^2=1$ and also we conclude that $2\mathrm{Re}\langle b,\Psi(z)\rangle=0$ for all $z$ on the neighborhood. Because $\Psi(z)$ is holomorphic, $\langle b|\Psi(z)\rangle$ is holomorphic. It is zero at the origin and its real part is always zero. Hence, we conclude that it is zero. It then follows that $\Phi(z)=U\Psi(z) + b$, where the two terms in the right-hand side are mutually orthogonal and $b$ is a unit vector. We can then arrange for a unitary $V\in\mathrm{U}(N+M)$ such that $\Phi(z)=V(1,\Psi_1(z),\dots, \Psi_{N-1}(z),0,0,\dots, 0)$. Multiplying this equation by $f_1(z)$, we find
\begin{align}
g(z)=V(f_1(z),\dots,f_N(z),0,\dots,0).    
\end{align}
This is valid on a neighbourhood of $z=0$. By using the power series expansion, we can extend this to any value of $z$. Due to the identity theorem for holomorphic functions, this is the required identity. The proof is complete. We remark that the theorem also holds for the infinite dimensional case, where we replace $\mathbb{C}^N$ by the space of square-summable sequences $\ell^2(\mathbb{N})$, with the proof requiring no changes. The proof presented here is an adaptation of those in Refs.\cite{lawson:1980,green:1978}
\section{Proof of Eq.~\eqref{eq:metricandnormalizations} of the main text}
\label{sec:}

Observe that, due to holomorphicity,
\begin{align}
\widetilde{h}_N= \frac{\partial ^2\log \bra{\widetilde{\Psi}_{N,\bm{\theta}}}\widetilde{\Psi}_{N,\bm{\theta}}\rangle}{\partial \theta \partial\bar{\theta}}=\frac{\bra{\frac{\partial \widetilde{\Psi}_{N,\bm{\theta}}}{\partial \theta}}\frac{\partial \widetilde{\Psi}_{N,\bm{\theta}}}{\partial \theta}\rangle \langle \widetilde{\Psi}_{N,\bm{\theta}}|\widetilde{\Psi}_{N,\bm{\theta}}\rangle - \langle \frac{\partial \widetilde{\Psi}_{N,\bm{\theta}}}{\partial \theta}|\widetilde{\Psi}_{N,\bm{\theta}}\rangle\langle \widetilde{\Psi}_{N,\bm{\theta}}|\frac{\partial \widetilde{\Psi}_{N,\bm{\theta}}}{\partial \theta}\rangle }{\bra{\widetilde{\Psi}_{N,\bm{\theta}}}\widetilde{\Psi}_{N,\bm{\theta}}\rangle^2},
\end{align}
hence all that remains to prove is
\begin{align}
\bra{\frac{\partial \widetilde{\Psi}_{N,\bm{\theta}}}{\partial \theta}}\frac{\partial \widetilde{\Psi}_{N,\bm{\theta}}}{\partial \theta}\rangle \langle \widetilde{\Psi}_{N,\bm{\theta}}|\widetilde{\Psi}_{N,\bm{\theta}}\rangle - \langle \frac{\partial \widetilde{\Psi}_{N,\bm{\theta}}}{\partial \theta}|\widetilde{\Psi}_{N,\bm{\theta}}\rangle\langle \widetilde{\Psi}_{N,\bm{\theta}}|\frac{\partial \widetilde{\Psi}_{N,\bm{\theta}}}{\partial \theta}\rangle =   \bra{\widetilde{\Psi}_{N+1,\bm{\theta}}}\widetilde{\Psi}_{N+1,\bm{\theta}}\rangle\bra{\widetilde{\Psi}_{N-1,\bm{\theta}}}\widetilde{\Psi}_{N-1,\bm{\theta}}\rangle. 
\end{align}
This can be shown as follows. We have that, 
\begin{align}
\frac{\partial}{\partial \theta}\ket{\widetilde{\Psi}_{N,\bm{\theta}}}= \ket{\widetilde{u}_{\bm{\theta}}}\wedge  \frac{\partial}{\partial \theta}\ket{\widetilde{u}_{\bm{\theta}}}\wedge \dots \wedge \frac{\partial ^{N-2}}{\partial \theta^{N-2}}\ket{\widetilde{u}_{\bm{\theta}}}\wedge \frac{\partial ^{N}}{\partial \theta^{N}}\ket{\widetilde{u}_{\bm{\theta}}} =\ket{\widetilde{\Psi}_{N-1,\bm{\theta}}}\wedge  \frac{\partial ^{N}}{\partial \theta^{N}}\ket{\widetilde{u}_{\bm{\theta}}}= \ket{\widetilde{\Psi}_{N-1,\bm{\theta}}}\wedge \left(\widetilde{Q}_{N-1}\frac{\partial ^{N}}{\partial \theta^{N}}\ket{\widetilde{u}_{\bm{\theta}}}\right),
\end{align}
where $\widetilde{Q}_N=1-\widetilde{P}_N$ and $\widetilde{P}_N$ being the orthogonal projector which is naturally associated to filling the $N$ first GLLs
\begin{align}
\widetilde{P}_N=\sum_{j=0}^{N-1}\ket{\widetilde{u}_{j,\bm{\theta}}}\bra{\widetilde{u}_{j,\bm{\theta}}}.    
\end{align}
It then follows that
\begin{align}
\bra{\frac{\partial \widetilde{\Psi}_{N,\bm{\theta}}}{\partial \theta}}\frac{\partial \widetilde{\Psi}_{N,\bm{\theta}}}{\partial \theta}\rangle= \langle \widetilde{\Psi}_{N-1,\bm{\theta}}|\widetilde{\Psi}_{N-1,\bm{\theta}}\rangle \bra{\frac{\partial ^{N}}{\partial \theta^{N}}\widetilde{u}_{\bm{\theta}}} Q_{N-1}\ket{\frac{\partial ^{N}}{\partial \theta^{N}}\widetilde{u}_{\bm{\theta}}}
\end{align}
and
\begin{align}
\bra{\frac{\partial \widetilde{\Psi}_{N,\bm{\theta}}}{\partial \theta}}\widetilde{\Psi}_{N,\bm{\theta}}\rangle= \langle \widetilde{\Psi}_{N-1,\bm{\theta}}|\widetilde{\Psi}_{N-1,\bm{\theta}}\rangle \bra{\frac{\partial ^{N}}{\partial \theta^{N}}\widetilde{u}_{\bm{\theta}}} \widetilde{Q}_{N-1}\ket{\frac{\partial ^{N-1}}{\partial \theta^{N-1}}\widetilde{u}_{\bm{\theta}}}
\end{align}
Also, since
\begin{align}
\ket{\widetilde{\Psi}_{N+1,\bm{\theta}}}=\ket{\widetilde{\Psi}_{N,\bm{\theta}}}\wedge \frac{\partial ^{N}}{\partial \theta^{N}}\ket{\widetilde{u}_{\bm{\theta}}}=\ket{\widetilde{\Psi}_{N,\bm{\theta}}}\wedge \left(\widetilde{Q}_N\frac{\partial ^{N}}{\partial \theta^{N}}\ket{\widetilde{u}_{\bm{\theta}}}\right),
\end{align}
it follows that
\begin{align}
\langle \widetilde{\Psi}_{N+1,\bm{\theta}}|\widetilde{\Psi}_{N+1,\bm{\theta}}\rangle= \langle \widetilde{\Psi}_{N,\bm{\theta}}|\widetilde{\Psi}_{N,\bm{\theta}}\rangle\bra{\frac{\partial ^{N}}{\partial \theta^{N}}\widetilde{u}_{\bm{\theta}}} \widetilde{Q}\ket{\frac{\partial ^{N}}{\partial \theta^{N}}\widetilde{u}_{\bm{\theta}}}.
\end{align}
Additionally, 
\begin{align}
\widetilde{P}_{N+1}=\widetilde{P}_{N} +\frac{\widetilde{Q}_{n-1}\frac{\partial ^{N}}{\partial \theta^{N}}\ket{\widetilde{u}_{\bm{\theta}}} \bra{\frac{\partial ^{N}}{\partial \theta^{N}}\widetilde{u}_{\bm{\theta}}}\widetilde{Q}_{N-1}}{\bra{\frac{\partial ^{N}}{\partial \theta^{N}}\widetilde{u}_{\bm{\theta}}} \widetilde{Q}_{N-1}\ket{\frac{\partial ^{N}}{\partial \theta^{N}}\widetilde{u}_{\bm{\theta}}}}
\end{align}
and, hence,
\begin{align}
\widetilde{Q}_{N+1}=\widetilde{Q}_{N} -\frac{\widetilde{Q}_{N-1}\frac{\partial ^{N}}{\partial \theta^{N}}\ket{\widetilde{u}_{\bm{\theta}}} \bra{\frac{\partial ^{N}}{\partial \theta^{N}}\widetilde{u}_{\bm{\theta}}}\widetilde{Q}_{N-1}}{\bra{\frac{\partial ^{N}}{\partial \theta^{N}}\widetilde{u}_{\bm{\theta}}} \widetilde{Q}_{N-1}\ket{\frac{\partial ^{N}}{\partial \theta^{N}}\widetilde{u}_{\bm{\theta}}}}.
\end{align}
Collecting everything, it then follows that
\begin{align}
&\bra{\frac{\partial \widetilde{\Psi}_{N,\bm{\theta}}}{\partial \theta}}\frac{\partial \widetilde{\Psi}_{N,\bm{\theta}}}{\partial \theta}\rangle \langle \widetilde{\Psi}_{N,\bm{\theta}}|\widetilde{\Psi}_{N,\bm{\theta}}\rangle - \langle \frac{\partial \widetilde{\Psi}_{N,\bm{\theta}}}{\partial \theta}|\widetilde{\Psi}_{N,\bm{\theta}}\rangle\langle \widetilde{\Psi}_{N,\bm{\theta}}|\frac{\partial \widetilde{\Psi}_{N,\bm{\theta}}}{\partial \theta}\rangle \nonumber \\
&=\langle \widetilde{\Psi}_{N-1,\bm{\theta}}|\widetilde{\Psi}_{N-1,\bm{\theta}}\rangle \bra{\frac{\partial ^{N}}{\partial \theta^{N}}\widetilde{u}_{\bm{\theta}}} \widetilde{Q}_{N-1}\ket{\frac{\partial ^{N}}{\partial \theta^{N}}\widetilde{u}_{\bm{\theta}}}\langle \widetilde{\Psi}_{N,\bm{\theta}}|\widetilde{\Psi}_{N,\bm{\theta}}\rangle \nonumber\\
&- \langle \widetilde{\Psi}_{N-1,\bm{\theta}}|\widetilde{\Psi}_{N-1,\bm{\theta}}\rangle ^2 \bra{\frac{\partial ^{N}}{\partial \theta^{N}}\widetilde{u}_{\bm{\theta}}} \widetilde{Q}_{N-1}\ket{\frac{\partial ^{N-1}}{\partial \theta^{N-1}}\widetilde{u}_{\bm{\theta}}}\bra{\frac{\partial ^{N-1}}{\partial \theta^{N-1}}\widetilde{u}_{\bm{\theta}}} \widetilde{Q}_{N-1}\ket{\frac{\partial ^{N}}{\partial \theta^{N}}\widetilde{u}_{\bm{\theta}}} \nonumber \\
&=\langle \widetilde{\Psi}_{N-1,\bm{\theta}}|\widetilde{\Psi}_{N-1,\bm{\theta}}\rangle \langle \widetilde{\Psi}_{N,\bm{\theta}}|\widetilde{\Psi}_{N,\bm{\theta}}\rangle \left(\bra{\frac{\partial ^{N}}{\partial \theta^{N}}\widetilde{u}_{\bm{\theta}}} \widetilde{Q}_{N-1}\ket{\frac{\partial ^{N}}{\partial \theta^{N}}\widetilde{u}_{\bm{\theta}}}-\frac{\bra{\frac{\partial ^{N}}{\partial \theta^{N}}\widetilde{u}_{\bm{\theta}}} \widetilde{Q}_{N-1}\ket{\frac{\partial ^{N-1}}{\partial \theta^{N-1}}\widetilde{u}_{\bm{\theta}}}\bra{\frac{\partial ^{N-1}}{\partial \theta^{N-1}}\widetilde{u}_{\bm{\theta}}} \widetilde{Q}_{N-1}\ket{\frac{\partial ^{N}}{\partial \theta^{N}}\widetilde{u}_{\bm{\theta}}}}{\bra{\frac{\partial ^{N-1}}{\partial \theta^{N-1}}\widetilde{u}_{\bm{\theta}}} \widetilde{Q}_{N-1}\ket{\frac{\partial ^{N-1}}{\partial \theta^{N-1}}\widetilde{u}_{\bm{\theta}}}}\right) \nonumber \\
&=\langle \widetilde{\Psi}_{N-1,\bm{\theta}}|\widetilde{\Psi}_{N-1,\bm{\theta}}\rangle \langle \widetilde{\Psi}_{N,\bm{\theta}}|\widetilde{\Psi}_{N,\bm{\theta}}\rangle \bra{\frac{\partial ^{N}}{\partial \theta^{N}}\widetilde{u}_{\bm{\theta}}} \widetilde{Q}_{N}\ket{\frac{\partial ^{N}}{\partial \theta^{N}}\widetilde{u}_{\bm{\theta}}}=\langle \widetilde{\Psi}_{N-1,\bm{\theta}}|\widetilde{\Psi}_{N-1,\bm{\theta}}\rangle \langle \widetilde{\Psi}_{N+1,\bm{\theta}}|\widetilde{\Psi}_{N+1,\bm{\theta}}\rangle,
\end{align}
as claimed.
\end{widetext}

\end{document}